\documentclass[12pt,a4paper]{article}
\begin{document}
\bibliographystyle{aip}
\newcommand{\nyk}{\mbox{\Large \{}}
\newcommand{\csk}{\mbox{\Large \}}}

\newcommand{\bra}[1]{\ensuremath{\langle #1 \vert}}
\newcommand{\ket}[1]{\ensuremath{\vert #1  \rangle}}
\newcommand{\braket}[2]{\ensuremath{\langle  #1\vert #2  \rangle}}
\newcommand{\ketbra}[1]{\ensuremath{\vert{#1}\rangle\langle{#1}\vert}}
\newcommand{\be}{\begin{equation}}
\newcommand{\ee}{\end{equation}}
\newcommand{\bea}{\begin{eqnarray}}
\newcommand{\eea}{\end{eqnarray}}
\newcommand{\vp}{\varphi}
\newcommand{\h}{\hat}
\newcommand{\ve}{\varepsilon}
\newcommand{\vrr}{\vec r}
\newcommand{\pr}{\;\prime}
\newcommand{\noi}{\noindent}
\newcommand{\vphi}{\varphi}
\newcommand{\ie}{{\it i.e., }}
\date{}

\markright{}
\title{Is the Spreading of Quantum Mechanical Wave Packets Indeed Inevitable? }
\author{Istv\'an Mayer\footnote{mayer@chemres.hu, mayer.istvan@ttk.mta.hu}\\ 
{\normalsize Research Centre for Natural
Sciences,}\\
{\normalsize Hungarian Academy of Sciences,
H-1525 Budapest, P.O.Box 17, Hungary}}

\setcounter{page}{0}
\maketitle
\begin{abstract}

It is demonstrated that -- contrary to the common belief -- it is possible
to construct solutions of the non-relativistic Schr\"odinger equation
of a free particle,
that do not exhibit dispersion. 
However, it seems that no  {\it normalizable\/} wave packets 
can be built up by their use, so the spreading of the wave packets is 
indeed inevitable.

\end{abstract}

\thispagestyle{empty}

\newpage

\subsection*{1. Introduction}

It is usually considered that the limiting transition from the quantum 
mechanical description to the classical mechanical one is not as 
straightforward as is the transition from the mechanics of the special theory 
of relativity to the classical mechanics. In fact, Ehrenfest theorem 
[\ref{Blokhincev}]
stating that the center of a wave packet moves according to Newton's laws
could be sufficient to follow a smooth transition from quantum mechanics
to the classical one as $\hbar \rightarrow 0$, but the picture is spoiled
by the observation that quantum mechanical wave packets are spreading in the
time until they fill the whole space. This is the case because it is usually
considered that the solutions of the Schr\"odinger equations of a free
(unbounded) particle always exhibit dispersion: the phase speed of
a de Broglie wave differs from the velocity of the particle it describes.


It is to be mentioned that non-spreading solutions have been described
in the literature, but not for free particles. Thus, 
such solutions are possible if there
is also an external potential. The well-known example is a Gaussian
wave packet in the potential well of a harmonic oscillator, but also
non-spreading wave-packets have been considered in the field of a microwave 
source, {\it etc} [\ref{kalinski}--\ref{buchleitner}]. 
Non-spreading wave packets 
can be constructed also
by adding a non-linear term to the Schr\"odinger equation [\ref{klein}].

A special word is appropriate about the one-dimensional case. Berry and Balazs
[\ref{BerBa}] showed that there is a sort of ``Airy-packets" that is a solution 
of the Schr\"odinger equation of a free particle and moves
without spreading---however, it is {\it accelerating\/} in a manner as if
there were a constant external potential. Later the uniquenes of this solution
was also proved [\ref{UnnRau}], which means that (except, of course,
the standart infinite sinusoidal waves) these accelerating Airy packets are the only
possible soulutions for an one-dimensional Shr\"odinger equation for a free 
particle that move  
without change of shape. Admitting three dimensions, however, some additional
mathematical freedom appears, too.

From the reasons mentioned above, it seems to be of interest to check whether 
the wave packets moving with a constant speed in the three-dimensional
free space are indeed {\it inevitably\/}
spreading out in the time. For that reason one has to consider the most general
wave packets that are non-spreading {\it by construction\/}.
%
%
%
%
It will be seen that by using cylindrical coordinates one can, indeed, 
construct solutions of the non-relativistic Schr\"odinger 
equation of a free particle 
that do not exhibit dispersion, but there appears an unsurmountable
{\it normalization problem\/} that prevents them to be used to build
up physically meaningful wave packets.


\subsection*{2. Cylindrical waves without dispersion}

Let us consider a free particle moving with the velocity $v$ along the positive direction of the
axis $z$ of a cylindrical system of coordinates $z,\, r,\, \vphi$. 
We shall search the wave function in the form
\be
\label{psi}
\Psi = R(r)f(z-vt)
\ee
This means that for sake 
of simplicity we shall assume the wave function $\Psi$ to be independent of
coordinate $\vphi$, so $\partial\Psi/\partial\vphi=0$. (Admitting a 
\mbox{$\vp$-dependence} would not be significant from our point of view; it has also
been shown that solutions with $\vp$-dependence can always be generated from
the $\vp$-independent ones in a trivial manner [\ref{leckner}].) 

Obviously, any function of form (\ref{psi}) represents a ``wave" moving with 
the phase-velocity $v$ in the positive direction of the axis $z$. This
function $\Psi$ is required to be a solution of the time-dependent
Schr\"odinger equation
\be
i\hbar\frac{\partial\Psi}{\partial t}=\hat H\Psi
\ee
As we are dealing with a free particle, the Hamiltonian is simply
\be
\h H = -\frac{\hbar^2}{2m}\Delta
\ee
where $m$ is the mass of the particle. Using the expression of the
Laplacian in the cylindrical coordinates selected, 
and taking into account that $\Psi$ is independent of the angle $\vphi$,
we get
\be
i\hbar R\frac{\partial f}{\partial t}=-\frac{\hbar^2}{2m}\left[
R\frac{\partial^2f}{\partial z^2} + f\frac1r\frac{\partial}{\partial r}
\left(r\frac{\partial R}{\partial r}\right)
\right]
\ee
By performing a standard separation of variables, and taking into account
that $f$ depends on $z$ and $t$ only through their combination $z-vt$, we get
the coupled system of ordinary differential equations
\bea
\label{syst}
f^{\,\prime\prime} - i\frac{2mv}{\hbar}f^{\,\prime}-qf=0\nonumber\\[-3mm]
\\[-3mm] \nonumber
\frac1r\frac{d}{dr}\left(r\frac{dR}{dr}\right) +qR=0
\eea
where $q$ is an arbitrary constant. (Also see bellow.) 

The general solution of the system (\ref{syst}) is, as easy to see,
\bea
f(u)&=& 
C_1exp\left[\frac1{\hbar}\left(imv+\sqrt{q\hbar^2-m^2v^2}\right)u\right]
\nonumber \\[-2mm] \\[-2mm]
&+& C_2exp\left[\frac1{\hbar}\left(imv-\sqrt{q\hbar^2-m^2v^2}\right)u\right]
\nonumber
\eea
where we have introduced the notation
\be
u=z-vt
\ee
and
\be
R(r)=C_3J_0(\sqrt{q}\!\;r) + C_4Y_0(\sqrt{q}\!\;r)
\ee
where $J_0$ and $Y_0$ are the standard cylindrical (Bessel) functions.

The solutions obtained remain finite everywhere if one requires
\be
0 \le q \le \frac{m^2v^2}{\hbar^2}
\ee
and sets 
\be
C_4=0
\ee

In the special case of selecting $q=0$ one has for $R(r)$ the special solution
\be
\left. R(r)\right|_{q=0} = C_3 + C_4ln(r)
\ee
which again remains finite only if $C_4=0$. However, in this case we
have simply $R(r)= Const$ and we recover the special case of the plane wave
which is constant in the directions orthogonal to $z$. In that case
\be
\label{plane}
f(u) = C_1 + C_2exp\left(\frac{2mvi}\hbar u\right)
\ee
thus -- disregarding the constant part, \ie by setting $C_1=0$ -- we get a 
standard exponential plane wave for which the phase velocity is half of the
group velocity [\ref{Blokhincev}] which equals the conventional velocity of the 
particle.\footnote{At the beginning we have selected the {\it phase velocity} 
to be equal $v$, therefore Eq.~(\ref{plane}) describes a de  Broglie wave 
corresponding to the conventional velocity $2v$.} 

Independently of the value of the parameters $q$ and $C_i$, all these solutions
for the Schr\"odinger equation of the free particle are simply shifted
by {\it the same\/} distance $vt$ along the positive axis $z$
after the period of time $t$, as they depend only on the combination
$u=z-vt$ but not on $z$ and $t$ separately. That means
that these functions
do not exhibit dispersion and any their linear combinations    
conserve their forms, only are shifted in the space as time passes. 
Therefore the wave packets formed as linear combinations of these functions
\bea
\label{lincomb}
\Psi(z,r,t)&=& 
\int_{0}^{q_{max}}\left\{A(q)exp\left[\frac{i}{\hbar}\left(mv+\sqrt{m^2v^2-q\hbar^2}\right)(z-vt)\right]\right.
 \\[2mm] && \left.
\quad + B(q)exp\left[\frac{i}{\hbar}\left(mv-\sqrt{m^2v^2-q\hbar^2}\right)(z-vt)\right]\right\}J_0(\sqrt{q}\!\;r)dq
\nonumber  
\eea
where $q_{max}=(mv/\hbar)^2$ and $A(q)$, $B(q)$ are arbitrary functions,
do not spread out in the space, as consist of components having {\it the same
phase velocity} $v$.
That is the fundamental difference with respect to the ``usual" wave packets
formed of the plane waves, the different components of which always have
{\it different phase velocities}, and therefore are inevitably 
spreading out. 

However, these wave packets are clearly not normalizable. Even the first 
inspection shows that the components coresponding to a given $q$
do not decay 
to zero as variable $z$ tends to $\pm \infty$. It is clear that
one simply cannot select the functions $A(q)$ and $B(q)$ as to provide 
a thorought cancellation of the terms at infinite intervals. In full
accord with this, by utilizing the
closure-relation of Bessel functions,
\be
\int_0^\infty xJ_\alpha(ux)J_\alpha(vx)dx= \frac1{u}\delta(u-v)\ , \quad  \quad
\alpha > -\frac1{2};
\ee
one obtains for the normalization integral of the wave function (\ref{lincomb}),
calculeted for t=0 (or any fixed value of $t$),
the expression 
\bea
\label{norm}
\left.
\langle \Psi|\Psi\rangle\right|_{t=0}&=& 2\pi
\int_{-\infty}^{\infty} dz \int_0^{q_{max}}\nyk |A(q)|^2 + |B(q)|^2 
\\ && \quad +
2Re\left[A(q)B^*(q)exp\left(\frac{2zi}{\hbar}\sqrt{m^2v^2-q\hbar^2}\right)
\right]\csk\frac{dq}{q} \nonumber
\eea
This is an integral over an infinite interval with a non-negative
integrand: the oscillating third terms of the integrand cannot overcompensate 
the positive first two -- at most, if $|A(q)|=|B(q)|$, it can just compensate 
them  in individual points.\footnote{That can be seen by subtracting and 
adding
$2|A(q)||B(q)|$. Then the first two terms of tne integrand give 
$\left(|A(q)|-|B(q)|\right)^2$ while the second becomes 
$|A(q)||B(q)|2Re\left\{1 + exp\left[i\left([(\psi_A(q)-\psi_B(q)]+
\frac{2z}{\hbar}\sqrt{m^2v^2-q\hbar^2}\right]\right)\right\}$, neither of which can 
be negative. [Here $\psi_A(q)$ and $\psi_B(q)$ are the phases of $A(q)$ and 
$B(q)$, respectively.]
} 
Therefore the normalization integral (\ref{norm})
diverges.

\subsection*{3. Conclusions}
It is demonstrated that using cylindrical coordinates one can form wave packets
out of different functions all of which have the same phase velocity
and all of them are solutions of the non-relativistic Schr\"odinger
equations of a free particle. Such wave packets do not spread out---but are not 
normalizable. Therefore the spreading of the physically relevant normalizable
wave packets is indeed inevitable.

\newpage

\subsection*{References}
\begin{enumerate}
\item
\label{Blokhincev}
D.~I. Blokhincev,
{\em Osnovy Kvantovoi Mekhaniki}, Vysshaia Shkola, Moscow, 1963.

\item
\label{kalinski}
M. Kalinski and J. H. Eberly,   Phys. Rev. {\it A} {\bf 53}, 1715 (1996)

\item
\label{arminjon}
M. Arminjon, Nuovo Cimento {\bf 114B}, 71 (1999)

\item
\label{buchleitner}
A. Buchleitner, D. Delande and J. Zakrzewski, Physics Reports {\bf 368}, 
409 (2002) 

\item
\label{klein}
A.G. Klein and S.A. Werner, Rep. Prog. Phys., {\bf 46}, 259 (1983)

\item
\label{BerBa}
M. V. Berry and N.L. Balazs, Am. J. Phys. {\bf 47}, 264 (1979)

\item
\label{UnnRau}
K. Unnikrishnan and A. R. P. Rau, Am. J. Phys. {\bf 64}, 1034 (1996)

\item
\label{leckner}
J. Lekner,  Eur. J. Phys. {\bf 29} 1121 (2008) 

\end{enumerate}
\end{document}